\documentclass{ifacconf}
\usepackage{graphicx}      
\usepackage{float}
\usepackage{amsmath, amssymb}
\usepackage{longtable} 
\usepackage{natbib}
\usepackage{comment}
\usepackage{psfrag}
\usepackage{epstopdf, epsfig}
\usepackage{xcolor}
\usepackage{subcaption} 
\begin{document}
\begin{frontmatter}
\title{ Dynamics of the Thermomagnetic Pendulum\thanksref{footnoteinfo}} 
\thanks[footnoteinfo]{This research was supported by the NASA Missouri Space Grant, Grant number: 80NSSC20M0100-2025\_AAC.}
\author[First]{Ryan Thompson} 
\author[Second]{Ethan Wang} 
\author[Third]{Nilay Kant}

\address[First]{Department of Mechanical and Aerospace Engineering, Missouri University of Science \& Technology, Rolla, Missouri, USA \\(e-mail: rjtb8p@mst.edu)}
\address[Second]{Department of Mechanical and Aerospace Engineering, Missouri University of Science \& Technology, Rolla, Missouri, USA \\(e-mail: eawnkc@mst.edu)}
\address[Third]{Department of Mechanical and Aerospace Engineering, Missouri University of Science \& Technology, Rolla, Missouri, USA \\(e-mail: nilaykant@mst.edu)}

\begin{abstract}
A thermomagnetic pendulum is introduced as a coupled thermo-magnetic-mechanical system consisting of a ferromagnetic bob under gravity and an offset permanent magnet. Heating drives the bob temperature above and below the Curie point, causing magnetic attraction to vanish and recover as the bob moves and cools. A multiphysics model is developed in which the magnetic torque depends nonlinearly on the bob temperature field and pendulum configuration. The formulation couples transient three-dimensional heat transfer, a temperature-dependent magnetization law, and pendulum dynamics. Simulations show angular torque asymmetry, rapid force reduction near the Curie point, and sustained oscillations.               
\end{abstract}
\end{frontmatter}

\section{Introduction}

Thermomagnetic energy conversion has long been investigated for low-grade thermal energy harvesting, with early pyromagnetic and thermomagnetic motor concepts described by \cite{Edison_1888_PyromagneticMotor} and \cite{Tesla_1889_ThermomagneticMotor}. Recent studies by \cite{dzekan2021efficient}, \cite{Joseph_thermomag_generators_2022}, and \cite{Liu_MagnetocaloricSwitchTMG_2023} have continued to examine thermomagnetic materials and devices for waste-heat recovery and related applications. \cite{kishore2018review} distinguish between \emph{active} thermomagnetic devices, which convert thermal input directly into electrical output through temperature-driven magnetic-flux variation, and \emph{passive} devices, which first generate motion through temperature-dependent magnetic interaction and only then convert that motion into useful output. Similar distinctions are used by \cite{Jiang_RegenerativeStaticTMG_2022} and \cite{Liu_ActiveTMG_2023}, while \cite{Christiaanse_Bruck_StaticTMG_2014} present a proof-of-concept experimental static thermomagnetic generator representative of the active class. On the passive side, recent examples include application-oriented generators studied by \cite{MorenoResendiz_MarineTMG_2025}, thermomagnetic and magnetic-piezoelectric harvesters examined by \cite{Ujihara_Carman_Lee_2007} and \cite{chen2015miniature}, coupled harvesting models developed by \cite{Bulgrin_Ju_Carman_Lavine_2009} and \cite{Joshi_Priya_Multiphysics_2013}, and thermomagnetic motor studies reported by \cite{deJesus_ThermodynamicEvaluationTMM_2024} and \cite{Silva_RotaryTMM_2025}.

The system considered in this paper is a passive thermomagnetic pendulum consisting of a ferromagnetic bob suspended as a rigid pendulum near an offset permanent magnet and subjected to localized heating over part of its motion. As discussed by \cite{Zeeshan_ThermomagneticHeatEngine_2021}, ferromagnetic response changes strongly near the Curie point. As the bob temperature rises, magnetization and magnetic attraction weaken; the bob then moves away from the heated region, cools, and regains magnetic attraction. The device thus uses its natural dynamics to sustain cyclic motion while limiting continued thermal exposure.

Modeling such a system is nontrivial because the magnetic torque cannot be prescribed from the mechanical equations alone. Instead, it emerges from the coupled thermal state of the bob, the temperature-dependent magnetic response of the material, and the geometry-dependent interaction between the bob and the external field. As shown by \cite{Song_RectangularMagnetPendulum_2023} and \cite{Song_MechanicalAntenna_Torque_2025}, predicted force and torque in magnetic pendulum-type systems depend sensitively on field representation, spatial field distribution, and geometry. Related thermomagnetic-device studies likewise show that computational treatment is needed when thermal and magnetic states evolve in time. For example, \cite{RodriguezMendez_TMEC_2D}, \cite{Joshi_Priya_Multiphysics_2013}, and \cite{Phillips_Carman_ActiveTMG_2020} each develop numerical thermomagnetic-device models, while \cite{Bulgrin_Ju_Carman_Lavine_2009} show the need for coupled thermal and mechanical treatment in a thermally driven harvesting device. Taken together, these studies indicate that the present problem cannot be reduced to a pendulum with an externally prescribed forcing law.

This paper develops a coupled thermo-magnetic-mechanical computational framework for a passive thermomagnetic pendulum. The model resolves the transient three dimensional thermal evolution of the bob, uses that temperature field to determine the local magnetic response, and evaluates the magnetic torque in the pendulum dynamics.

\section{System Description and Problem Formulation}

\begin{figure}[b!]
 \centering
\psfrag{A}[c][c][1][0]{\small $\gamma$}
\psfrag{B}[c][c][1][0]{\small $\ell$}
\psfrag{C}[c][c][1][0]{\small $\theta$}
\psfrag{D}[c][c][1][0]{\small $\phi$}
\psfrag{E}[c][c][1][0]{\small $R$}
\psfrag{F}[c][c][1][0]{\small $g$}
\psfrag{G}[c][c][1][0]{\small $L_{\rm gap}$}
\psfrag{i}[c][c][1][0]{\small  $x$}
\psfrag{j}[c][c][1][0]{\small  $y$}
\psfrag{k}[c][c][1][0]{\small  $z$}
\psfrag{v}[c][c][1][0]{\small  $r$}
\psfrag{W}[c][c][1][0]{\small $O'$}
\psfrag{H}[c][c][1][0]{\small {uniform heat source}}

 \includegraphics[width=0.65\linewidth]{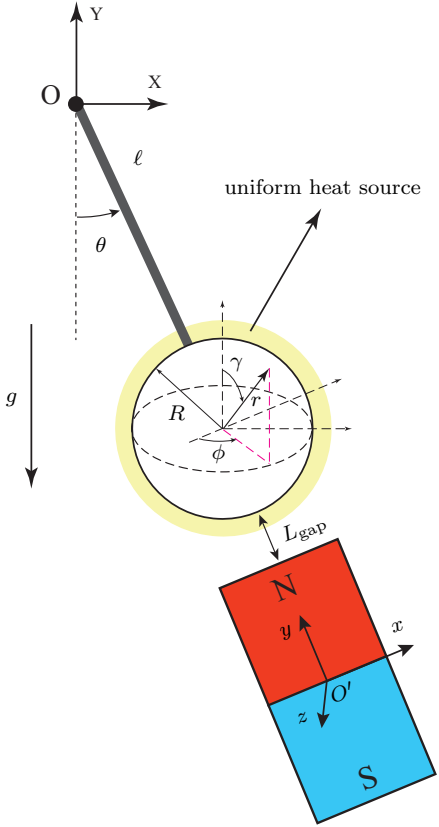}
\caption{The thermomagnetic pendulum.}
 \label{Fig1}
 \end{figure}

Consider the thermomagnetic pendulum shown in Fig.~\ref{Fig1}. The system consists of a rigid pendulum of length $\ell$ with a ferromagnetic spherical bob of radius $R$ attached at its distal end, and a permanent magnet placed at an offset from the lowest vertical position. When the bob and magnet are aligned, the minimum separation distance is denoted by $L_{\rm gap}$. The motion is assumed planar, and the pendulum angle $\theta$ is measured counterclockwise from the downward vertical. A global reference frame is fixed at the pivot $O$, while local frames are attached to the bob and the magnet, as shown in Fig.~\ref{Fig1}$.$

The operating principle is based on the Curie effect, as described by \cite{Zeeshan_ThermomagneticHeatEngine_2021}, \emph{i.e.} the magnetization of a ferromagnetic material depends on temperature and rapidly decreases to as the temperature approaches the Curie point $T_c$. The bob is subjected to uniform heating over a prescribed angular region, leading to a temperature field that evolves with the pendulum position and velocity. As a result, the magnetic interaction between the bob and the permanent magnet becomes temperature dependent through the local magnetization. At temperatures below $T_c$, the bob remains magnetized and experiences a magnetic force toward the magnet. As the temperature increases toward and beyond $T_c$, the magnetic interaction weakens, and the motion becomes dominated by gravity. As the bob exits the heated region, it cools through convection and radiation, and magnetization is recovered once the temperature falls below $T_c$. The magnetic interaction is then re-established, producing a restoring torque that drives the return motion toward the magnet. This cyclic heating and cooling process induces oscillatory motion.

Under these effects, the pendulum dynamics is given by
\begin{equation}
I\,\ddot{\theta} + c\,\dot{\theta} + mgL\sin\theta + \tau_{\mathrm{mag}}(\theta,T) = 0
\label{eq:pendulum_dynamics}
\end{equation}

\noindent where $I$ is the mass moment of inertia about $O$, $c$ is the viscous damping coefficient at the pivot, $m$ is the bob mass, $g$ is the gravitational acceleration, and $L \triangleq \ell + R$ is the distance from $O$ to the bob center. The term $\tau_{\mathrm{mag}}(\theta,T)$ denotes the magnetic torque, which depends on both the pendulum configuration and the bob temperature.

It is noteworthy that $\tau_{mag}$ cannot be determined from the mechanical model alone. Its evaluation requires a coupled multiphysics description involving three interacting subsystems: a thermal subsystem that predicts the transient temperature field within the bob, a magnetic subsystem that computes the resulting temperature-dependent bob-magnet interaction, and a mechanical subsystem that advances the pendulum motion under the induced torque. Developing a coupled thermal-magnetic-mechanical model of the system dynamics is the main objective of this paper.


\section{Thermal Subsystem Modeling}
Since $\tau_{mag}$ in \eqref{eq:pendulum_dynamics} depends on the bob temperature, heat transfer within the spherical bob is modeled as transient conduction with no internal volumetric heat generation. The governing heat equation in spherical coordinates is given by \cite{Incropera2011}:
\begin{equation}\label{eq:heat-equation}
\begin{aligned}
\rho C_p \frac{\partial T}{\partial t}
&=
\frac{1}{r^2}\frac{\partial}{\partial r}
\left( k r^2 \frac{\partial T}{\partial r} \right)
+
\frac{1}{r^2 \sin\gamma}\frac{\partial}{\partial \gamma}
\left( k \sin\gamma \frac{\partial T}{\partial \gamma} \right) \\[0.5em]
&\quad +
\frac{1}{r^2 \sin^2\gamma}\frac{\partial}{\partial \phi}
\left( k \frac{\partial T}{\partial \phi} \right)
\end{aligned}
\end{equation}

\noindent where \(T = T(r,\gamma,\phi,t)\) is the temperature, $\rho$ is the density, $C_p$ is the specific heat capacity, and $k$ is the thermal conductivity.  Note that the thermo-physical properties $\rho$, $C_p$ and $k$ are assumed constant. At the outer surface $r = R$, the heat flux balance accounts for applied heating, convection, and radiation, which leads to the following boundary conditions seen in \cite{Incropera2011}:
\begin{align}\label{eq:boundary-conditions}
-k \left.\frac{\partial T}{\partial r}\right|_{r=R}
&=
q_{\mathrm{ext}}(R,\gamma,\phi,t)
+
h(v)\bigl(T(R,\gamma,\phi,t)-T_\infty\bigr)
\\[-0.7em]
&\quad +
\epsilon \sigma \bigl(T(R,\gamma,\phi,t)^4 - T_\infty^4\bigr) \notag
\end{align}

where $q_{\mathrm{ext}}$ denotes the prescribed surface heating, $h(v)$ is the convective heat transfer coefficient, $T_\infty$ is the ambient temperature, $\epsilon$ is the emissivity, and $\sigma$ is the Stefan-Boltzmann constant. Although the externally applied heat flux is uniform over the surface of the bob, the convective cooling is nonuniform; therefore, no geometric symmetry is assumed in the temperature distribution. 

The convective coefficient $h$ in \eqref{eq:boundary-conditions} depends on the instantaneous surface velocity $v$, thereby coupling the thermal and mechanical subsystems. It is evaluated using standard forced-convection correlations for flow over a sphere:
\begin{equation}
h(v) = \frac{Nu\,k_\infty}{D}, \qquad D = 2R
\label{eq:conv_boundary}
\end{equation}
where the Nusselt number $Nu$ is given by \cite{RanzMarshall1952}:

\[
Nu =
\begin{cases}
2 + 0.6\,Re^{1/2}\,Pr^{1/3}, & Re < 200, \\
0.37\,Re^{0.6}\,Pr^{1/3}, & 200 < Re < 10^5
\label{eq:Nusselt}
\end{cases}
\]

\noindent In the expression above, $Re$ and $Pr$ denote the Reynolds and Prandtl numbers that obey:
\begin{align}
Re(v) = \frac{\rho_f\,v\,D}{\mu_f}, \qquad
Pr = \frac{\mu_f\,C_{pf}}{k_\infty}
\label{eq:Re_Pr} 
\end{align}

\noindent Here, $\rho_f$, $\mu_f$, $C_{pf}$, and $k_\infty$ denote the fluid density, dynamic viscosity, specific heat, and thermal conductivity, respectively. Furthermore, natural convection is not modeled separately and is assumed secondary except near low-velocity turning points. For the present smooth sphere and low-to-moderate Reynolds number regime, wake effects are neglected, and the convective coefficient is evaluated from the local incident flow, consistent with the assumptions in \cite{White2011} and \cite{Incropera2011}. These relations couple the thermal and mechanical subsystems through the instantaneous bob velocity, and the resulting temperature field provides the input to the magnetization model in Section 4.


\section{Magnetic Subsystem Modeling}

\subsection{Magnetostatic Framework and Effective Field}

To evaluate the magnetic contribution $\tau_{\mathrm{mag}}$ in \eqref{eq:pendulum_dynamics}, the bob-magnet interaction is modeled under magnetostatic conditions, neglecting displacement currents and induction effects. In the absence of free currents, the governing equations reduce to
\begin{equation}
\nabla \cdot \mathbf{B} = 0, \qquad \nabla \times \mathbf{H} = 0
\end{equation}
with constitutive relation
\begin{equation}
\mathbf{B} = \mu_0 (\mathbf{H} + \mathbf{M})
\end{equation}
where $\mathbf{B}$ denotes the magnetic flux density, $\mathbf{H}$ the magnetic field intensity, $\mathbf{M}$ the magnetization, and $\mu_0$ the permeability of free space, as shown in \cite{Fernow_Magnetostatics_2016}.

The magnetic field acting on the ferromagnetic bob is generated by an external permanent magnet. In the present reduced model, the external flux density $\mathbf{B}_{\mathrm{ext}}(\mathbf{r})$ is computed analytically using a uniformly magnetized cuboid model developed by \cite{EngelHerbert_StrayField_2005}, and the corresponding applied field in the bob region is taken as $\mathbf{H}_{\mathrm{ext}}(\mathbf{r}) = \mathbf{B}_{\mathrm{ext}}(\mathbf{r})/\mu_0$.

\subsection{External Magnetic Field Representation}

The external field generated by the permanent magnet is evaluated analytically using closed-form expressions for a uniformly magnetized cuboid \cite{EngelHerbert_StrayField_2005}. Let $(x_i,y_i,z_i)$ denote the coordinates of the evaluation point relative to the magnet center, and let $L_x$, $L_y$, and $L_z$ denote the half-length, half-width, and half-height of the magnet. Let $(M_x,M_y,M_z)$ denote the components of the uniform magnetization. The external magnetic flux density at $\mathbf{p}$ is given by
\begin{equation}
\mathbf{B}_\mathrm{ext}(\mathbf{p})
=
\frac{\mu_0}{4\pi}
\sum_{s_1=\pm1}\sum_{s_2=\pm1}\sum_{s_3=\pm1}
s_1 s_2 s_3
\begin{bmatrix}
B_x \\ B_y \\ B_z
\end{bmatrix}
\label{eq:engel_field}
\end{equation}
where
\[
\begin{aligned}
B_x &= M_x \arctan\!\frac{y_i - s_2 L_y}{(x_i - s_1 L_x) G}
      - M_y \log\!\big(G - z_i + s_3 L_z\big) \\
    &\quad + M_z \log\!\big(G - y_i + s_2 L_y\big) \\[0.5ex]
B_y &= M_x \log\!\big(G - z_i + s_3 L_z\big) \\
    &\quad - M_y \arctan\!\left(
      \frac{(x_i - s_1 L_x)(z_i - s_3 L_z)}
           {(y_i - s_2 L_y)\,G}
      \right) \\
    &\quad - M_z \log\!\big(G - x_i + s_1 L_x\big) \\[0.5ex]
B_z &= - M_x \log\!\big(G - y_i + s_2 L_y\big)
      + M_y \log\!\big(G - x_i + s_1 L_x\big) \\
    &\quad + M_z \arctan\!\frac{(x_i - s_1 L_x)(y_i - s_2 L_y)}{(z_i - s_3 L_z) G}
\end{aligned}
\]
with
\[
G \triangleq \sqrt{(x_i - s_1 L_x)^2 + (y_i - s_2 L_y)^2 + (z_i - s_3 L_z)^2}
\]

This external field interacts with the spatially varying magnetization within the bob. Its spatial gradients generate magnetic body forces, which in turn contribute to the torque $\tau_{\mathrm{mag}}$ in \eqref{eq:pendulum_dynamics}.

Within the ferromagnetic bob, the internal field is reduced by demagnetizing effects induced by the self-generated magnetization. In the present reduced-order model, these effects are approximated using the spherical demagnetization factor $N = 1/3$, yielding the effective field
\begin{equation}
\mathbf{H}_\mathrm{eff}(\mathbf{r})
=
\mathbf{H}_\mathrm{ext}(\mathbf{r})
- N \mathbf{M}(\mathbf{r})
\end{equation}

This relation provides a mean-field approximation of the internal magnetic response, capturing the leading-order geometric attenuation of the applied field without requiring a full magnetostatic boundary-value solution.

\subsection{Temperature-Dependent Nonlinear Magnetization}

Given the effective field defined above, the bob's local magnetization must be modeled as a temperature-dependent constitutive response. This is because the thermal field computed in Section~3 drives the material through a temperature range where its magnetic behavior changes rapidly. For nickel, this transition is dominated by the loss of ferromagnetic order as the temperature approaches $T_c$. The operating regime is therefore restricted to temperatures near $T_c$, where this effect governs the magnetic response.

The saturation magnetization is modeled using a critical scaling approximation described by \cite{nagalakshmi2020}:  
\begin{equation}
M_s(T) =
\begin{cases}
M_{s0} \left( 1 - \dfrac{T}{T_c} \right)^\beta, & T < T_c \\
0, & T \ge T_c
\end{cases}
\label{eq:sat_mag}
\end{equation}
where $M_{s0}$ is the zero-temperature saturation magnetization, $\beta$ is a material-dependent critical exponent, and $T$ is the absolute temperature. 

For $T < T_c$, the magnetization is assumed to align with the effective field, while its magnitude follows a smooth nonlinear saturation law. Specifically, a hyperbolic tangent approximation to a Langevin-type response is used:
\begin{equation}
\mathbf{M}(\mathbf{r}) =
M_s(T)\tanh\!\left(
\frac{|\mathbf{H}_\mathrm{eff}(\mathbf{r})|}{M_s(T)}
\right)\hat{\mathbf{H}}_\mathrm{eff}(\mathbf{r}),
\quad T < T_c
\label{eq:tanh_approx}
\end{equation}

\noindent where $\hat{\mathbf{H}}_\mathrm{eff} = \mathbf{H}_\mathrm{eff}/|\mathbf{H}_\mathrm{eff}|$. For $T \ge T_c$, the magnetization vanishes, i.e., $\mathbf{M}(\mathbf{r}) = \mathbf{0}$. The resulting spatially varying magnetization field is then used to compute the distributed magnetic forces as a function of temperature and subsequently, the resulting torque $\tau_{\mathrm{mag}}$; this is presented next.

\subsection{Magnetic Force and Torque}

The force of attraction between the magnetized bob and the external field is modeled through a magnetic body-force density. For a differential volume element, the force density is taken as
\begin{equation}
\mathbf{f}(\mathbf{r}) = \bigl(\mathbf{M}(\mathbf{r}) \cdot \nabla\bigr)\mathbf{B}_{\mathrm{ext}}(\mathbf{r}),
\qquad
\mathbf{B}_{\mathrm{ext}}(\mathbf{r}) = \mu_0 \mathbf{H}_{\mathrm{ext}}(\mathbf{r}),
\end{equation}
which captures the action of the nonuniform applied field on the spatially varying magnetization within the bob. The resulting magnetic force and the corresponding torque about the pivot are obtained by integrating this body-force density over the bob volume, consistent with how \cite{Meeker_FEMM} approaches standard computational treatments of magnetic force and torque evaluation:
\begin{equation}\label{eq:torque}
\mathbf{F} = \int_V \mathbf{f}(\mathbf{r})\, dV,
\qquad
\boldsymbol{\tau} = \int_V \bigl(\mathbf{L} + \mathbf{r}\bigr) \times \mathbf{f}(\mathbf{r})\, dV
\end{equation}

\noindent where $\mathbf{L}$ denotes the vector from the pivot $O$ to the bob center, $\mathbf{r}$ is the local position vector relative to the bob center (see Fig. \ref{Fig1}), and $V$ is the bob volume. The net magnetic torque $\tau_\mathrm{mag}$ in \eqref{eq:pendulum_dynamics} is finally given by $\boldsymbol{\tau}$, acting along the axis passing through the pivot $O$. It should be noted that magnetic force and torque computations in \eqref{eq:torque} neglect electromagnetic transients and hysteresis. 

\section{Numerical Implementation of the Coupled Model}

This section describes the numerical implementation of the coupled thermal, magnetic, and mechanical model developed in Sections~3 and~4. At each time step, the transient temperature field within the bob is first computed, the resulting temperature-dependent magnetization is then updated, and $\tau_{\mathrm{mag}}$ is evaluated from the corresponding force distribution. The computed torque is subsequently used to simulate the dynamics in \eqref{eq:pendulum_dynamics}.

\subsection{Computational Mesh}

The bob is discretized using a structured spherical grid in $(r,\gamma,\phi)$, with node-centered control volumes defined by midpoint interfaces between adjacent grid points. This yields a vertex-centered finite-volume formulation, with connectivity implicitly defined through neighboring nodes in each coordinate direction. All fields are defined on the same mesh. In particular, the temperature $T(r,\gamma,\phi)$ and magnetization $\mathbf{M}(r,\gamma,\phi)$ are co-located at identical control volumes, ensuring that thermo-magnetic coupling is evaluated locally without interpolation.

\subsection{Thermal Discretization and Time Integration}

The thermal field is discretized on the structured spherical mesh defined in $(r,\gamma,\phi)$ using a node-centered finite volume formulation. Starting from the governing heat equation in \eqref{eq:heat-equation}, the radial, polar, and azimuthal diffusion terms are approximated separately on each control volume, yielding the discrete directional operators $D_r$, $D_\gamma$, and $D_\phi$. These operators represent the local conductive contributions from neighboring nodes in each coordinate direction. For each control volume, they are written as:

\begin{align} \label{eq:D_terms}
D_r 
&=
\frac{k}{r^2 (\Delta r)^2}
\Bigl[
\Bigl(r+\tfrac{\Delta r}{2}\Bigr)^2
\bigl(T_{r + \Delta r,\gamma,\phi}^{n+1} - T_{r,\gamma,\phi}^{n+1}\bigr)
 \\
&\quad
-
\Bigl(r-\tfrac{\Delta r}{2}\Bigr)^2
\bigl(T_{r,\gamma,\phi}^{n+1} - T_{r - \Delta r,\gamma,\phi}^{n+1}\bigr)
\Bigr]
\cr
D_\gamma 
&=
\frac{k}{r^2 \sin\gamma (\Delta\gamma)^2}
\Bigl[
\sin\!\Bigl(\gamma + \tfrac{\Delta \gamma}{2}\Bigr)
\bigl(T_{r,\gamma + \Delta \gamma,\phi}^{n+1} - T_{r,\gamma,\phi}^{n+1}\bigr)
\nonumber \cr
&\quad
-
\sin\!\Bigl(\gamma - \tfrac{\Delta \gamma}{2}\Bigr)
\bigl(T_{r,\gamma,\phi}^{n+1} - T_{r,\gamma - \Delta \gamma,\phi}^{n+1}\bigr)
\Bigr]\cr
D_\phi 
&=
\frac{k}{r^2 \sin^2\gamma (\Delta\phi)^2}
\Bigl(
T_{r,\gamma,\phi + \Delta \phi}^{n+1}
- 2T_{r,\gamma,\phi}^{n+1}
+ T_{r,\gamma,\phi - \Delta \phi}^{n+1}
\Bigr)
\label{eq:D_terms}
\end{align}

These discrete operators define a local stencil coupling each control volume to its immediate neighbors. Assembly over the full domain yields a global system for the temperature field at the next time step $n+1$. Since the spatial operators depend on $T^{n+1}$, the formulation is fully implicit and requires the solution of a coupled system across all control volumes at each time step.


The temporal evolution of the temperature field is obtained through a first-order update of the form
\begin{equation}
T_{r,\gamma,\phi}^{n+1}
=
T_{r,\gamma,\phi}^{n}
+
\frac{\Delta t}{\rho C_p}
\left(
D_r + D_\gamma + D_\phi
\right)
\end{equation}

which represents the rearranged finite-volume discretization of~\eqref{eq:heat-equation} under an implicit time integration scheme, where the spatial operators are evaluated using the fully discretized finite-volume formulation in~\eqref{eq:D_terms}. 

\subsection{Magnetization Update and Torque Evaluation}

The magnetic response is evaluated on the same computational mesh, with the magnetization at each control volume determined by the nonlinear constitutive relation defined in Section~4. Since the effective field depends on the magnetization through the demagnetization model, the magnetization is updated iteratively at each time step until consistency is achieved between the constitutive relation and the effective field. Once the magnetization field has converged, the magnetic body-force density is evaluated and integrated over the bob volume to obtain the corresponding torque about the pivot, as defined in Section~4. This torque is then utilized in \eqref{eq:pendulum_dynamics} to simulate the pendulum dynamics.

\section{Results}

\begin{figure}[t!]
 \centering
\psfrag{A}[c][c][1][0]{\small $L_{\rm gap} = 1 \,\rm{cm}$}
\psfrag{B}[c][c][1][0]{\small $L_{\rm gap} = 1.5 \, \rm{cm}$}
\psfrag{C}[c][c][1][0]{\small $L_{\rm gap} = 2 \, \rm{cm}$}
\psfrag{D}[c][c][1][0]{\small $\mathbf{F}$ (N) vs $T$ ($^{\circ}\rm{C}$)}
\psfrag{E}[c][c][1][0]{\small $T = T_c$}
 \includegraphics[width=0.94\linewidth]{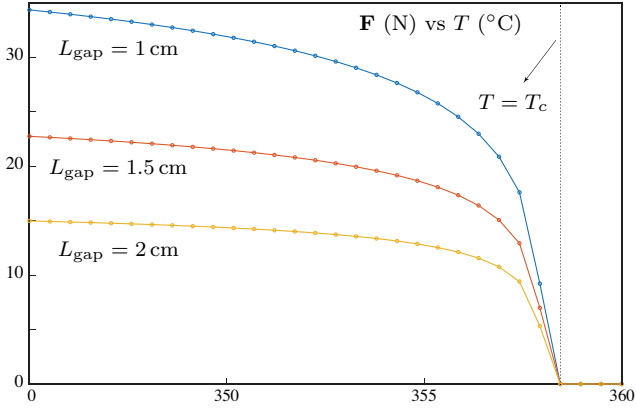}
\caption{Magnetic force vs. temperature for multiple gap lengths $L_{\mathrm{gap}}$ at $\theta = 30^\circ$.}
 \label{Fig2}
 \end{figure}

\begin{figure}[t!]
 \centering
\psfrag{A}[c][c][1][0]{\small $L_{\rm gap} = 1 \,\rm{cm}$}
\psfrag{B}[c][c][1][0]{\small $L_{\rm gap} = 1.5 \, \rm{cm}$}
\psfrag{C}[c][c][1][0]{\small $L_{\rm gap} = 2 \, \rm{cm}$}
\psfrag{D}[c][c][1][0]{\small $\tau_{\rm mag}$ (Nm) vs $\theta$ (deg.)}
 \includegraphics[width=0.94\linewidth]{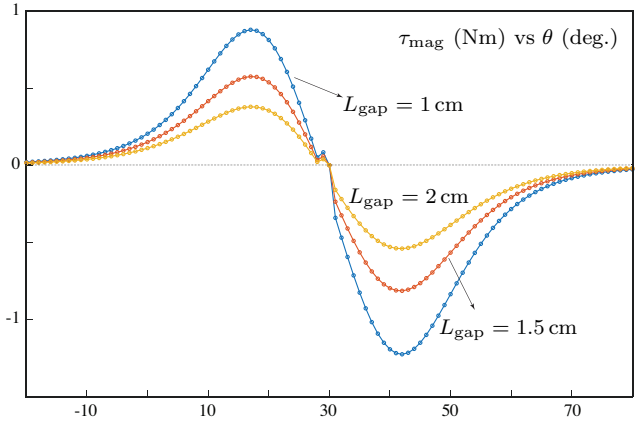}
\caption{Magnetic torque as a function of $\theta$ for multiple gap lengths $L_{\mathrm{gap}}$, evaluated at $T = 345^\circ\mathrm{C}$.}
 \label{Fig3}
 \end{figure}



The thermomagnetic model is evaluated using parametric sweeps and a time-domain simulation. Fig.~2 presents the magnetic force as a function of temperature at $\theta=30^\circ$ for three values of the separation distance $L_{\rm gap}$ using parameters in Table~1. Fig.~3 presents the magnetic torque as a function of pendulum angle for a uniform temperature of $T=345^\circ\mathrm{C}$, evaluated for the same $L_{\rm gap}$ values and parameters in Table~1. Figs.~4(a)-(b) show the angular displacement and angular velocity of the pendulum using the parameters in Tables~1 and~2.

Fig.~2 shows the temperature dependence of the magnetic force under a uniform temperature field, corresponding to the temperature-dependent magnetization model in \eqref{eq:sat_mag}-\eqref{eq:tanh_approx}. The force decreases nonlinearly with temperature, with a gradual reduction at lower temperatures followed by a rapid collapse as the Curie point $T_c$ is approached, and it becomes zero for $T \ge T_c$. This behavior follows directly from $M_s(T)$ in \eqref{eq:sat_mag}, since the magnetic force scales with the temperature-dependent magnetization, which decreases to zero as $T \to T_c$. Increasing $L_{\mathrm{gap}}$ reduces the force magnitude across all temperatures due to weaker field strength and reduced spatial gradients, while preserving the same functional dependence on temperature. 

Fig.~3 shows the angular dependence of $\tau_{\mathrm{mag}}$ in \eqref{eq:pendulum_dynamics} for a uniform, constant-temperature case. The torque is zero at $\theta=30^\circ$, where the net magnetic force produces no moment about the pivot. The positive and negative extrema are asymmetric due to the $30^\circ$ orientation of the permanent magnet (Fig.~1), which changes the alignment between the bob and the magnet pole face. For $\theta > 30^\circ$, the bob is more aligned with the pole face and produces a stronger torque contribution, whereas for $\theta < 30^\circ$ the alignment is weaker, reducing the effective moment. The torque magnitude also decreases with increasing $L_{\mathrm{gap}}$, consistent with reduced field strength and spatial gradients in \eqref{eq:engel_field}. A small localized feature near $\theta=29^\circ$, more accurately viewed as a reduction near $\theta=28^\circ$, arises from stronger edge-region gradients in the finite-sized magnet model \eqref{eq:engel_field}, which locally perturb the force distribution while preserving the overall torque trend.

\begin{figure}[b!]
 \centering
\psfrag{A}[c][c][1][0]{\small $\theta$ (deg.)}
\psfrag{B}[c][c][1][0]{\small $\dot \theta$ (rad/s)}
\psfrag{E}[c][c][1][0]{\small time (sec)}
 \includegraphics[width=0.95\linewidth]{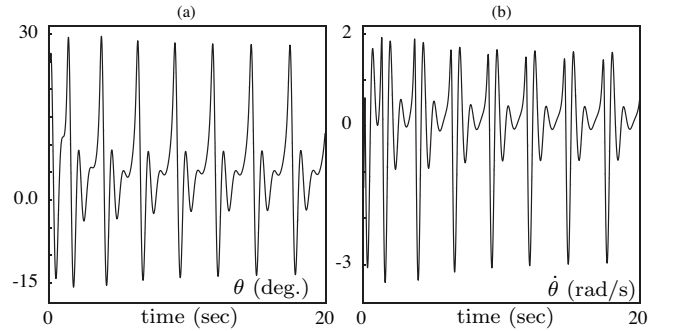}
\caption{(a) Angular displacement $\theta$ and (b) angular velocity $\dot{\theta}$ as functions of time.}
 \label{Fig4}
 \end{figure}
 
\begin{table}[t!]
\caption{Parameters for Parametric Sweeps}
\centering
\begin{tabular}{l c c}
\hline
\textbf{Parameter} & \textbf{Value} & \textbf{Units} \\
\hline
Pendulum bob material & Nickel & — \\ 
Magnet side length & 0.05 & m \\
Magnet remanent flux density & 1.2 & T \\
Magnet orientation angle & 30 & deg \\
Pendulum rod length & 0.1 & m \\
Pendulum bob radius & 0.015 & m \\
\hline
\end{tabular}
\end{table}

\begin{table}[t!]
\caption{Parameters for Time Simulation}
\centering
\begin{tabular}{l c c}
\hline
\textbf{Parameter} & \textbf{Value} & \textbf{Units} \\
\hline
Magnet–pendulum gap & 0.01 & m \\
Nickel density & 8765 & kg/m$^3$ \\
Nickel heat capacity & 555 & J/(kg$\cdot$K) \\
Nickel thermal conductivity & 65 & W/(m$\cdot$K) \\
Ambient temperature & 25 & $^\circ C$ \\
Initial bob temperature & 357.8 & $^\circ C$ \\
Gravity & 9.81 & m/s$^2$ \\
Damping coefficient & 0.005 & kg$\cdot$ m$^2$ / s \\
Pendulum bob mass & 0.124 & kg \\
Simulation time step & 0.001 & s \\
Total simulation time & 20 & s \\
Heating power & 50 & W \\
Surface emissivity & 0.8 & — \\
Stefan--Boltzmann constant & $5.67 \times 10^{-8}$ & W/(m$^2$K$^4$) \\
Initial angle & 30 & deg \\
Initial angular velocity & 0 & rad/s \\
\hline
\end{tabular}
\end{table}

Figs.~4(a)-(b) show the time evolution, governed by \eqref{eq:pendulum_dynamics}. As the bob enters the heated region, its temperature increases, reducing the saturation magnetization via \eqref{eq:sat_mag}-\eqref{eq:tanh_approx} and decreasing the magnetic torque, which drives motion away from the magnet. As the bob exits the heated region, it cools according to the boundary conditions in \eqref{eq:boundary-conditions}, restoring magnetization and increasing the magnetic torque. The resulting dependence of $\tau_{\mathrm{mag}}(\theta, T)$ on the coupled mechanical and thermal variables produces sustained oscillations and converges to a steady-state periodic motion.

\section{Conclusion}

A coupled thermo-magnetic-mechanical model of a pendulum with a ferromagnetic bob is developed. The formulation combines transient three-dimensional heat conduction, a nonlinear magnetization law with demagnetization effects, and rigid-body pendulum dynamics. The temperature field is computed using a node-centered finite-volume discretization on a structured spherical mesh and used to update the local magnetization, from which the magnetic torque is obtained. Simulations show that the response depends strongly on pendulum angle, bob temperature, and bob-magnet gap. The magnetic torque varies nonlinearly with angle and exhibits pronounced asymmetry due to the geometry and orientation of the applied field. Increasing separation reduces the torque magnitude through attenuation of the field and its spatial gradients, while increasing temperature weakens the magnetic response through reduction of the saturation magnetization, which vanishes at the Curie point. This temperature dependence leads to alternating loss and recovery of torque, producing sustained steady-state oscillatory motion. Future work will consider more general three-dimensional geometries, improved surface heat-transfer models, refined magnetic constitutive laws, experimental validation, and stability of the resulting periodic motion.


\bibliographystyle{plain}
\bibliography{ref}

\begin{thebibliography}{28}
\providecommand{\natexlab}[1]{#1}
\providecommand{\url}[1]{\texttt{#1}}
\providecommand{\urlprefix}{URL }
\expandafter\ifx\csname urlstyle\endcsname\relax
  \providecommand{\doi}[1]{doi:\discretionary{}{}{}#1}\else
  \providecommand{\doi}{doi:\discretionary{}{}{}\begingroup
  \urlstyle{rm}\Url}\fi

\bibitem[{Bulgrin et~al.(2009)Bulgrin, Ju, Carman, and
  Lavine}]{Bulgrin_Ju_Carman_Lavine_2009}
Bulgrin, K.E., Ju, Y.S., Carman, G.P., and Lavine, A.S. (2009).
\newblock A coupled thermal and mechanical model of a thermal energy harvesting
  device.
\newblock In \emph{Proceedings of the ASME 2009 International Mechanical
  Engineering Congress \& Exposition}.

\bibitem[{Chen et~al.(2015)Chen, Chung, Tseng, Hung, Yeh, and
  Cheng}]{chen2015miniature}
Chen, C.C., Chung, T.K., Tseng, C.Y., Hung, C.F., Yeh, P.C., and Cheng, C.C.
  (2015).
\newblock A miniature magnetic-piezoelectric thermal energy harvester.
\newblock \emph{IEEE Transactions on Magnetics}, 51(7), 1--9.

\bibitem[{Christiaanse and Br{\"u}ck(2014)}]{Christiaanse_Bruck_StaticTMG_2014}
Christiaanse, T. and Br{\"u}ck, E. (2014).
\newblock Proof-of-concept static thermomagnetic generator experimental device.
\newblock \emph{Metallurgical and Materials Transactions E}, 1(1), 36--40.

\bibitem[{de~Jesus et~al.(2024)de~Jesus, Santos, Silva, C{\^a}mara, Michel,
  Braga, Rowe, and Trevizoli}]{deJesus_ThermodynamicEvaluationTMM_2024}
de~Jesus, V.S., Santos, M.U.L., Silva, C.E.L., C{\^a}mara, M.A., Michel,
  H.C.C., Braga, C.M.P., Rowe, A., and Trevizoli, P.V. (2024).
\newblock Thermodynamic evaluation of thermomagnetic motors with first and
  second order transition magnetocaloric materials.
\newblock \emph{Applied Thermal Engineering}, 253, 123737.

\bibitem[{Dzekan et~al.(2021)Dzekan, Waske, Nielsch, and
  F{\"a}hler}]{dzekan2021efficient}
Dzekan, D., Waske, A., Nielsch, K., and F{\"a}hler, S. (2021).
\newblock Efficient and affordable thermomagnetic materials for harvesting low
  grade waste heat.
\newblock \emph{APL Materials}, 9(1).

\bibitem[{Edison(1888)}]{Edison_1888_PyromagneticMotor}
Edison, T.A. (1888).
\newblock Pyromagnetic motor.
\newblock U.S. Patent, patented Mar. 27, 1888.

\bibitem[{Engel-Herbert and Hesjedal(2005)}]{EngelHerbert_StrayField_2005}
Engel-Herbert, R. and Hesjedal, T. (2005).
\newblock Calculation of the magnetic stray field of a uniaxial magnetic
  domain.
\newblock \emph{Journal of Applied Physics}, 97(7).

\bibitem[{Fernow(2016)}]{Fernow_Magnetostatics_2016}
Fernow, R.C. (2016).
\newblock \emph{Principles of Magnetostatics}.
\newblock Cambridge University Press.

\bibitem[{Incropera et~al.(2011)Incropera, DeWitt, Bergman, and
  Lavine}]{Incropera2011}
Incropera, F.P., DeWitt, D.P., Bergman, T.L., and Lavine, A.S. (2011).
\newblock \emph{Fundamentals of Heat and Mass Transfer}.
\newblock John Wiley \& Sons, 7 edition.

\bibitem[{Jiang et~al.(2022)Jiang, Zhu, Yu, Luo, and
  Li}]{Jiang_RegenerativeStaticTMG_2022}
Jiang, C., Zhu, S., Yu, G., Luo, E., and Li, K. (2022).
\newblock Numerical and experimental investigations on a regenerative static
  thermomagnetic generator for low-grade thermal energy recovery.
\newblock \emph{Applied Energy}, 311, 118585.

\bibitem[{Joseph et~al.(2022)Joseph, Ohtsuka, Miki, and
  Kohl}]{Joseph_thermomag_generators_2022}
Joseph, J., Ohtsuka, M., Miki, H., and Kohl, M. (2022).
\newblock Thermal processes of miniature thermomagnetic generators in resonant
  self-actuation mode.
\newblock \emph{iScience}, 25(7), 104569.

\bibitem[{Joshi and Priya(2013)}]{Joshi_Priya_Multiphysics_2013}
Joshi, K.B. and Priya, S. (2013).
\newblock Multi-physics model of a thermo-magnetic energy harvester.
\newblock \emph{Smart Materials and Structures}, 22(5), 055005.

\bibitem[{Kishore and Priya(2018)}]{kishore2018review}
Kishore, R.A. and Priya, S. (2018).
\newblock A review on design and performance of thermomagnetic devices.
\newblock \emph{Renewable and Sustainable Energy Reviews}, 81, 33--44.

\bibitem[{Liu et~al.(2023{\natexlab{a}})Liu, Chen, Huang, Qiao, Yu, Xie,
  Ramanujan, Hu, Chu, Long et~al.}]{Liu_MagnetocaloricSwitchTMG_2023}
Liu, X., Chen, H., Huang, J., Qiao, K., Yu, Z., Xie, L., Ramanujan, R.V., Hu,
  F., Chu, K., Long, Y., et~al. (2023{\natexlab{a}}).
\newblock High-performance thermomagnetic generator controlled by a
  magnetocaloric switch.
\newblock \emph{Nature Communications}, 14(1), 4811.

\bibitem[{Liu et~al.(2023{\natexlab{b}})Liu, Zhang, Chen, Ma, Qiao, Xie, Ou,
  Wang, Hu, and Shen}]{Liu_ActiveTMG_2023}
Liu, X., Zhang, H., Chen, H., Ma, Z., Qiao, K., Xie, L., Ou, Z., Wang, J., Hu,
  F., and Shen, B. (2023{\natexlab{b}}).
\newblock Significant optimization of active thermomagnetic generator for
  low-grade waste heat recovery.
\newblock \emph{Applied Thermal Engineering}, 221, 119827.

\bibitem[{Meeker(2003)}]{Meeker_FEMM}
Meeker, D. (2003).
\newblock \emph{Finite Element Method Magnetics, Version 3.3, User's Manual}.
\newblock May 24, 2003.

\bibitem[{Mehmood and Cho(2021)}]{Zeeshan_ThermomagneticHeatEngine_2021}
Mehmood, M.U. and Cho, S. (2021).
\newblock Optimization of a thermomagnetic heat engine for harvesting low grade
  thermal energy.
\newblock \emph{Energies}, 14(18), 5768.

\bibitem[{Moreno~Resendiz et~al.(2025)Moreno~Resendiz, Peterson, and
  Kishore}]{MorenoResendiz_MarineTMG_2025}
Moreno~Resendiz, E., Peterson, T., and Kishore, R.A. (2025).
\newblock Thermomagnetic generators for ultra-low-grade marine thermal energy
  harvesting.
\newblock \emph{Communications Engineering}, 4(1), 204.

\bibitem[{Nagalakshmi et~al.(2020)}]{nagalakshmi2020}
Nagalakshmi, R. et~al. (2020).
\newblock Magnetocaloric effect and critical exponent studies in mn\_4. 5ni\_0.
  5sn\_3 alloy.
\newblock \emph{Journal of Chennai Academy of Sciences}, 2(2), 1--20.

\bibitem[{Phillips and Carman(2020)}]{Phillips_Carman_ActiveTMG_2020}
Phillips, M.R. and Carman, G.P. (2020).
\newblock Numerical analysis of an active thermomagnetic device for thermal
  energy harvesting.
\newblock \emph{Journal of Energy Resources Technology}, 142(8), 082102.

\bibitem[{Ranz and Marshall(1952)}]{RanzMarshall1952}
Ranz, W.E. and Marshall, W.R. (1952).
\newblock Evaporation from drops.
\newblock \emph{Chemical Engineering Progress}, 48(3), 141--146.

\bibitem[{Rodr{\'\i}guez-M{\'e}ndez et~al.(2024)Rodr{\'\i}guez-M{\'e}ndez,
  Gallo, Cugini, Fabbrici, Albertini, and Chin{\`e}}]{RodriguezMendez_TMEC_2D}
Rodr{\'\i}guez-M{\'e}ndez, F., Gallo, L., Cugini, F., Fabbrici, S., Albertini,
  F., and Chin{\`e}, B. (2024).
\newblock A 2d computational model of a thermomagnetic device.
\newblock In \emph{2024 Comsol Conference}.

\bibitem[{Silva et~al.(2025)Silva, Torres, C{\^a}mara, Michel, Braga, and
  Trevizoli}]{Silva_RotaryTMM_2025}
Silva, C.E.L., Torres, D.L.B., C{\^a}mara, M.A., Michel, H.C.C., Braga, C.M.P.,
  and Trevizoli, P.V. (2025).
\newblock Novel rotary thermomagnetic motor with a fin rotor for energy
  harvesting: Initial results.
\newblock \emph{Energy Conversion and Management}, 342, 120061.

\bibitem[{Song et~al.(2025)Song, Shuai, and
  Cheng}]{Song_MechanicalAntenna_Torque_2025}
Song, W., Shuai, C., and Cheng, Y. (2025).
\newblock Dynamic and magnetic field model of the magnetic pendulum mechanical
  antenna based on torque analysis.
\newblock \emph{Scientific Reports}, 15(1), 18475.

\bibitem[{Song et~al.(2023)Song, Shuai, and
  Zhai}]{Song_RectangularMagnetPendulum_2023}
Song, W., Shuai, C., and Zhai, Q. (2023).
\newblock A magnetic field model of rectangular permanent magnets in a magnetic
  pendulum array.
\newblock \emph{AIP Advances}, 13(10).

\bibitem[{Tesla(1889)}]{Tesla_1889_ThermomagneticMotor}
Tesla, N. (1889).
\newblock Thermo-magnetic motor.
\newblock U.S. Patent, filed Mar. 30, 1886, issued Jan. 15, 1889.

\bibitem[{Ujihara et~al.(2007)Ujihara, Carman, and
  Lee}]{Ujihara_Carman_Lee_2007}
Ujihara, M., Carman, G.P., and Lee, D.G. (2007).
\newblock Thermal energy harvesting device using ferromagnetic materials.
\newblock \emph{Applied Physics Letters}, 91, 093508.

\bibitem[{White(2011)}]{White2011}
White, F.M. (2011).
\newblock \emph{Fluid Mechanics}.
\newblock McGraw-Hill Education, 7 edition.

\end{thebibliography}

\end{document}